\documentclass[]{spie}  

\usepackage{amsmath,amsfonts,amssymb}
\usepackage{graphicx}
\usepackage[colorlinks=true, allcolors=blue]{hyperref}
\usepackage{xspace}
\usepackage{orcidlink}
\usepackage[utf8]{inputenc}
\usepackage[T1]{fontenc}

\title{An upgraded 0.4-meter telescope fleet for Las Cumbres Observatory's Educational and Science Programs}

\author[a]{Daniel-Rolf Harbeck\orcidlink{0000-0002-8590-007X}}
\author[a]{Brook Taylor}
\author[a]{Annie Kirby}
\author[a]{Mark Bowman\orcidlink{0000-0002-9716-6175}}
\author[a]{Steve Foale}
\author[a]{Kal Kadlec}
\author[a]{Curtis McCully\orcidlink{0000-0001-5807-7893}}
\author[a]{Matthew Daily}
\author[a]{Jon DeVera}
\author[a]{Dave Douglass}
\author[a]{Mark Willis}
\author[a]{Ian Baker}
\author[a]{Nikolaus Volgenau\orcidlink{0000-0002-0329-293X}}
\author[a]{Patrick Conway}
\author[a]{Brian Haworth}
\author[a]{Jesus Estrada}
\author[a]{Edward Gomez\orcidlink{0000-0001-5749-1507}}
\author[a]{Sandy Seale}
\author[a]{Alice Hopkinson}
\author[a]{Fernando Rios}
\author[a]{Prerana Kotapali}
\author[a]{Lisa Storrie-Lombardi\orcidlink{0000-0002-5987-5210}}
\author[a]{Wayne Rosing}
\affil[a]{Las Cumbres Observatory, Goleta, CA, USA}

\newcommand{\fits}{FITS\xspace}
\newcommand{\lco}{LCOGT\xspace}

\newcommand{\qhy}{QHY600\xspace}

\newcommand{\planewave}{PlaneWave\xspace}
\newcommand{\electron}{e$^-$\xspace}
\newcommand{\deltarho}{DeltaRho 350\xspace}
\newcommand{\haleakala}{Haleakal{\=a}\xspace}
\begin{document} 
\maketitle

\begin{abstract}

  Las Cumbres Observatory (\lco) operates a global network of robotic 0.4, 1.0, and 2.0-meter telescopes to facilitate scientific research and education in time-domain astronomy. 
  \lco's flagship educational program, Global Sky Partners (GSP), awards up to 1500 hours per year of telescope time to individuals and organizations that run their own, fully supported, educational programs. 
  The GSP has a presence in 40 countries and 45\% of the Partners target under-served, under-represented, and developing world audiences. 
  
  The degradation and obsolescence of the original 0.4-meter telescope network prompted \lco to update the fleet of 10 telescopes to a new system consisting of predominantly off-the-shelf products.
  New \planewave \deltarho telescopes with Gemini Focuser/Rotators, \lco filter wheels, and QHY600 CMOS cameras, complement the original, custom-built mount.
  The deployment of all ten telescopes was completed in March 2024. 
  
  We describe the design and performance of this new system and its components. 
  We comment on modifications made to the \qhy cameras, as well as on the treatment of random telegraph noise of its CMOS detectors within our data processing system BANZAI. 
  The new telescope network supports the GSP program as well as multiple key science projects, including follow-up observations for the TESS satellite mission. 
  
\end{abstract}


\section{A global telescope education network}
\label{sec:intro}  

\begin{figure}
    \centering
    \includegraphics[width=0.8\textwidth]{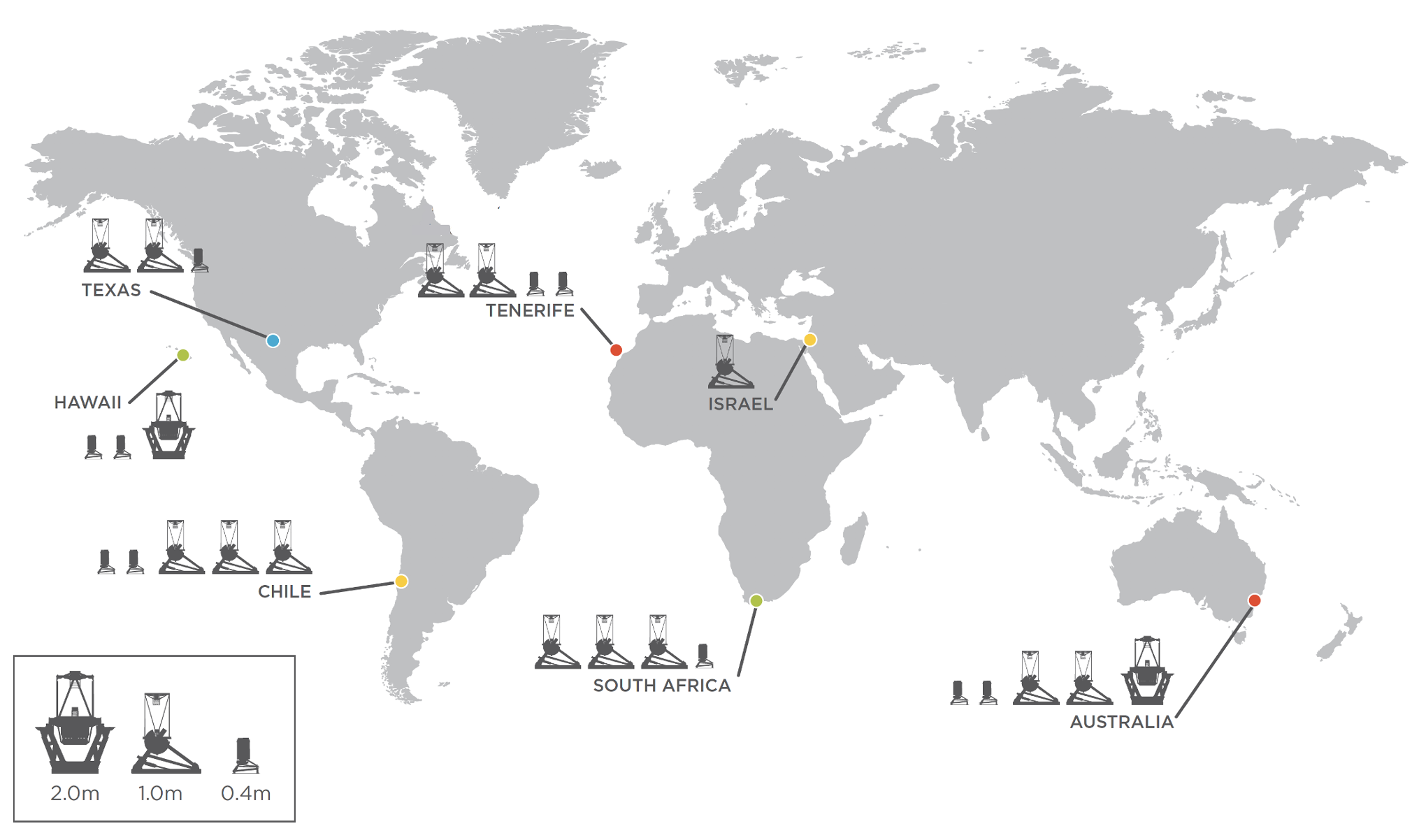}
    \caption{Las Cumbres Observatory's global telescope network consists of ten 0.4m telescopes, thirteen 1-meter telescopes, and two 2-meter telescopes at established observatory sites.}
    \label{fig:map}
\end{figure}

The mission of Las Cumbres Observatory\cite{brown_cumbres_2013} (\lco) is to “advance our understanding of the universe through science and education with our unique global telescope network.”
This {\em global telescope network} comprises ten 0.4-meter telescopes, thirteen 1-meter telescopes, and two 2-meter telescopes, see Figure~\ref{fig:map} for their locations.
\lco maintains an education program, the Global Sky Partners (GSP), that supports 30-35 educational groups around the world.
\lco provides the GSP with 1500 hours/year of telescope time, as well as web-based resources and a discussion forum. 
GSP participants run their own projects and can request observations from anywhere in the world; all they need is access to the internet.  
The primary purpose of \lco's network of the ten 0.4-meter telescopes is to provide observations for the GSP.
\lco science programs also routinely use this network to observe brighter targets. 
Two of \lco's current (2023-2026) Key Projects have been allocated hundreds of hours on the 0.4-meter telescopes to follow-up TESS transiting planet candidates and monitor oscillations in massive stars. 

The original (2016) 0.4-meter telescope network enabled \lco to establish the now far-reaching GSP program. The telescopes themselves were built on custom mounts and incorporated heavily modified Meade tubes, \lco-built filter wheels, and SBIG 6303 CCD cameras with integrated shutters. 
However, the productivity of the telescopes was lowered by the unreliability of the secondary mirror focus mechanism.
Too often, observations that were out-of-focus had to be repeated. 
To fix the focus mechanisms would have required unsupportable personnel resources, and success was not guaranteed. 
The SBIG cameras had reached the end of support by their vendor (now Diffraction Limited), and after multiple condensation events in the detector chambers over several years of operations, the electronic boards and detectors started to degrade (e.g., non-linear readout behavior) or outright fail.
As it was, operating the 0.4-meter network was no longer sustainable.

To rejuvenate the 0.4-meter network, we initiated a project to replace the Meade telescope tubes and the SBIG 6303 cameras at all sites.
As a first step, a single prototype telescope and instrument package were deployed to the \haleakala site in June 2022.
A select set of users from the GSP and key science projects were offered the opportunity to use the prototype.
The first production unit was deployed to McDonald Observatory in March 2023.
The final production units were deployed to CTIO in March 2024.
In all, ten telescopes were upgraded in the span of 13 months.
As shown in Figure~\ref{fig:map}, the telescopes are located at the following sites: \haleakala Observatory (Hawaii, USA), McDonald Observatory (Texas, USA), Teide Observatory (Spain), Cerro Tololo Inter-American Observatory (Chile) South African Astronomical Observatory (Sutherland, South Africa), and Siding Spring Observatory (Australia).

\section{PlaneWave DeltaRho 350 with a QHY600 camera}

From our experience with deploying and operating our global telescope network, we knew that an upgrade would pose both logistical (coordinating personnel for installation trips to six global sites) and fiscal challenges (procurements, assembly, and installation costs for ten units)  compared to a single telescope project.
The selected components described in the following reflect those trade-offs. 
To address those challenges, we made the decision to reuse existing and proven observatory and telescope infrastructure (both hardware and software) whenever feasible.
This decision guided us to preserve the C-mounts and filter wheels when we replaced the Meade telescopes.
The replacement we chose was the \planewave \deltarho telescope\footnote{https://planewave.com/product/deltarho350/}, which was mechanically compatible with the existing mounts, and thus allowed us to the reuse the existing telescope control hardware and software. 
The reduction in the aperture of the replacement telescopes was an acceptable cost for the (expected) benefit of improved overall performance that the refurbished system would provide.
The secondary mirror position is now fixed and the focus is achieved by translating the instrument package.

To replace the SBIG 6303 cameras, we chose the \qhy cameras, which have a pixel size that matches the plate scale of the \deltarho telescope.
The size ($36 \times  24$mm) of the QHY600 detector (Sony IMX 455) is well-matched to the unvignetted field of view, which in our case is limited by the 50 mm diameter filters in the filter wheels. 
The QHY600 is a CMOS-based camera with electronic shuttering.
Although this shuttering is perfect for nighttime observations, we decided to install an external (Uniblitz default-open) shutter to reject stray light when biases and darks are acquired during daytime. 
For focusing of the instrumentation package we chose a Gemini focuser/rotator as it proved reliable in our early testing of the prototype deployment.  
The rotator function is not used and remains unpowered in order to prevent unintended movement and subsequent mechanical interference of the instrumentation package (filter wheel, shutter, and camera) with the telescope mount.
All images from the \deltarho telescopes have a fixed orientation (North is up, and East is left) regardless of the hemisphere for the convenience of the users.

\begin{figure}
    \centering
    \includegraphics[width=0.9\textwidth]{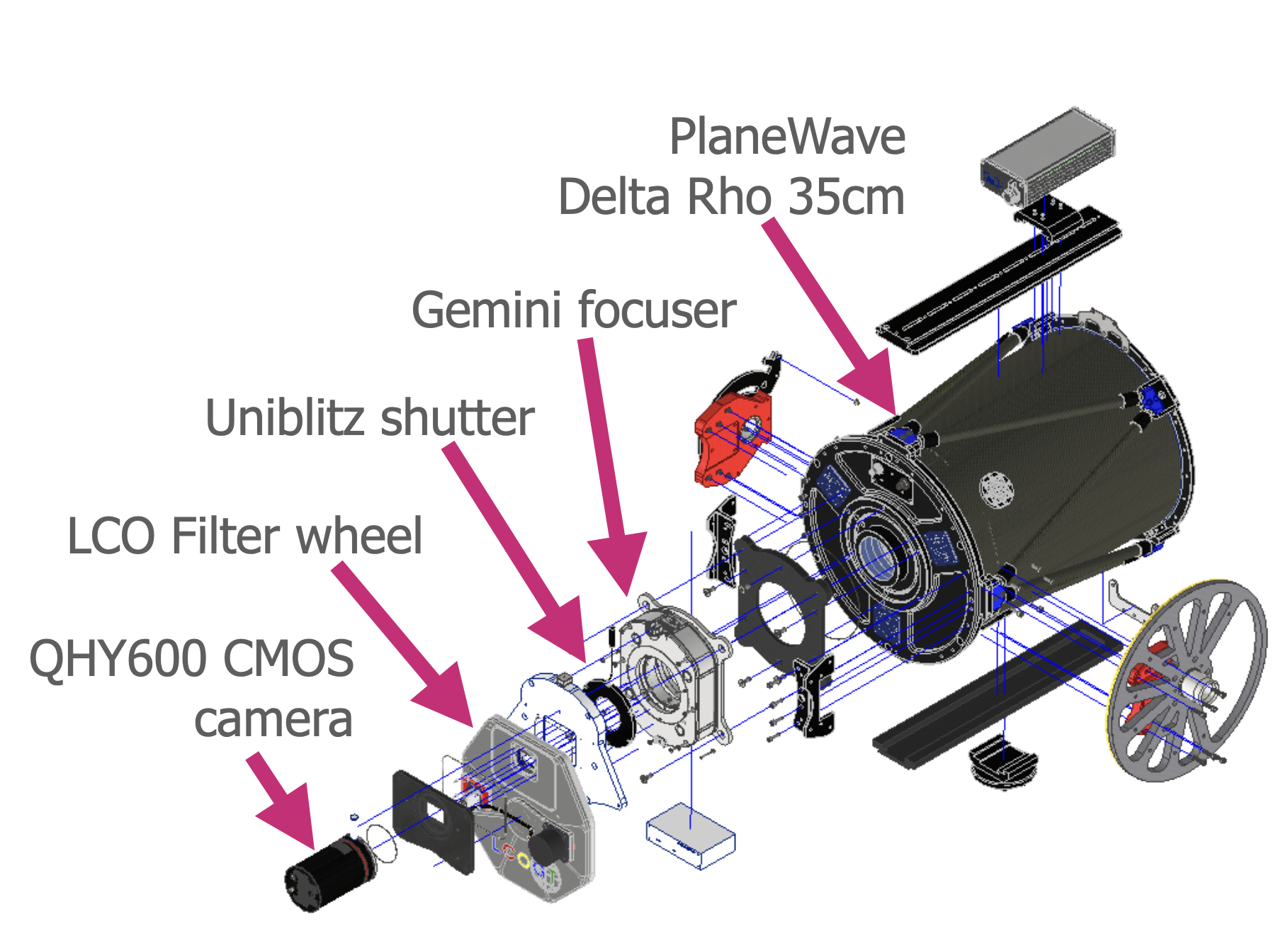}
    \vspace{1ex}
    \caption{Exploded view of the telescope and instrumentation system. }
    \label{fig:exploded-view}
\end{figure}

\begin{figure}
    \centering
    \includegraphics[width=0.9\textwidth]{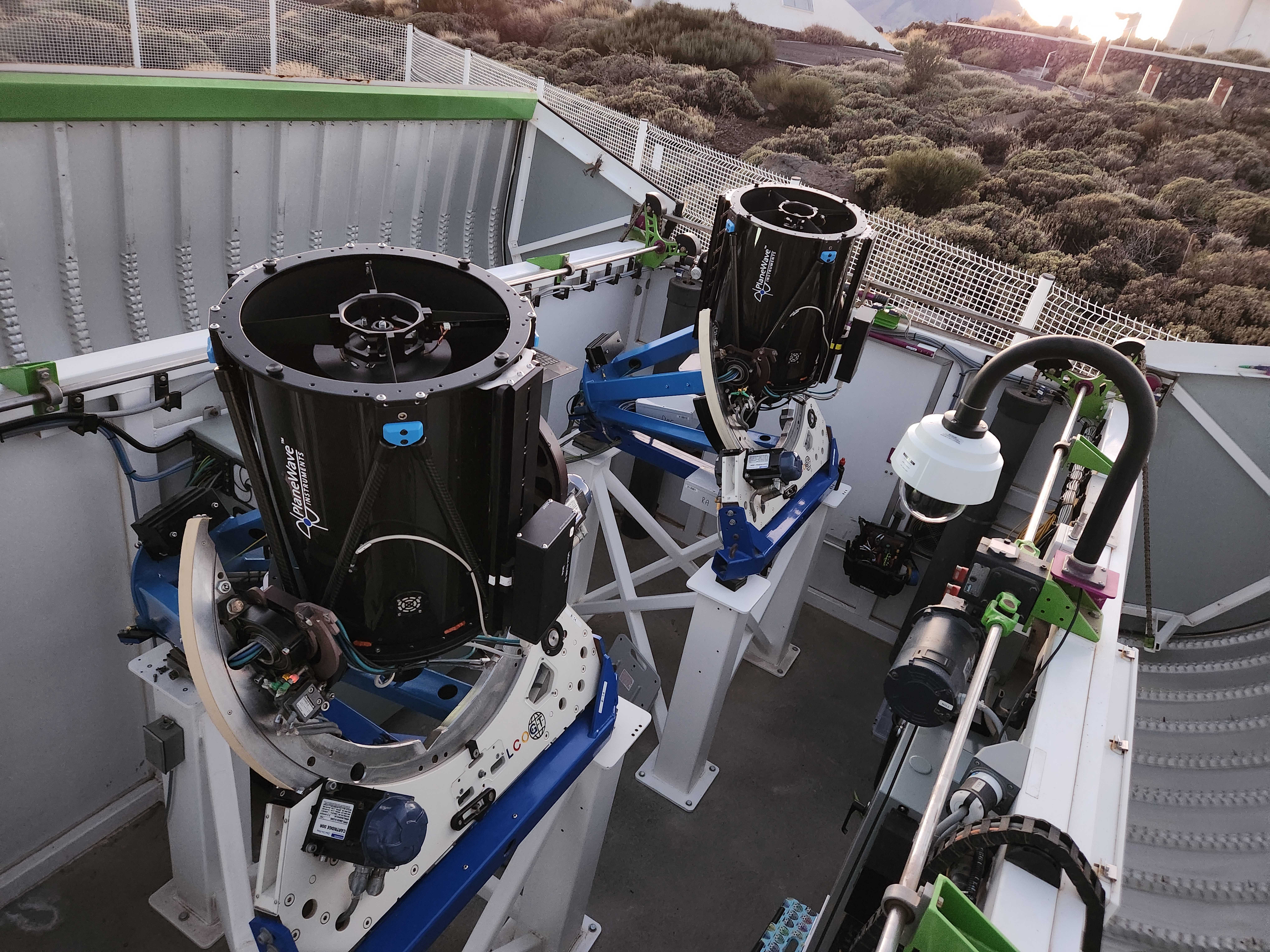}
    \vspace{1ex}
    \caption{Two PlaneWave \deltarho telescopes in the aqawan enclosure at Teide Observatory. }
    \label{fig:tfn}
\end{figure}

An exploded system view is shown in Figure \ref{fig:exploded-view}, and the deployment at the Teide Observatory site is shown in Figure \ref{fig:tfn}; the system's key parameters are summarized in Table \ref{tab:sysoverview}.

\subsection{Deployment}

The deployment at a site consisted of upgrading the electrical panel for the telescope to accommodate power and signal distribution, including upgraded USB3-compatible fiber extenders. 
The old instrumentation package and the 0.4-meter Meade telescopes are removed and, except for the filter wheels, donated or disposed of at the site. 
The filter wheels are sent back to the headquarters of \lco in Santa Barbara, CA, for refurbishment and integration into the next upgraded telescope system.

The installation process takes about 2-3 days per telescope and is followed by collimating the telescope on sky as per \planewave's procedure and shimming the instrument package to make the detector plane parallel to the focal plane. 
In the final step, super calibration data are collected, such as CMOS noise map, bias, dark, and flat field frames, but also telescope pointing data.
This can take up to a week, depending on weather conditions. 
In most cases, the new telescope was available for operation within two weeks after the start of the installation campaign.

\subsection{\deltarho Telescope}

The \planewave \deltarho telescope is a turn-key wide-field telescope with a primary aperture of 350mm.
The beam of the telescope is fast at f/3, which requires consideration when specifying the narrow-band filters. 

The \deltarho optical tube assembly (OTA) was field hardened for robotic operation prior to being deployed.
This included modifying the OTA cooling filters to a lower porosity filter media and enabling field replacement of the filter media by on-site technicians. Additional custom mounting hardware was fabricated to enable fine adjustment of OTA positions in the \lco mount.

During initial testing, low harmonic vibration was discovered to be exacerbated by the radial fans. Coupling the OTA tube with the end flanges via structural adhesion proved to mitigate this fan induced harmonic motion.

\subsection {\qhy camera and modifications}

The QHY600 camera, or more specifically, the Sony IMX455 CMOS light sensor, was chosen for the balance of cost efficiency and overall performance for the upgraded 0.4m telescope fleet.  
These cameras are commonly used in amateur astronomy but are also increasingly used in professional astronomy settings, especially in the planet transit follow-up community. 
An excellent description of these detectors is provided in Alarcon et al. (2023)\cite{alarcon_scientific_2023}. 
Our characterization of the QHY600 camera resulted in similar results and we generally find that the cameras operate within the vendor's specifications.  
Among the plethora of readout modes and gain settings available for the \qhy camera we choose a compromise setting ("High Gain Mode", Gain setting = 0) that enables both precision planet transit observations (priority on full well) and low surface brightness photometry (low read noise).
For specific use cases, better operating parameters may be established, however, \lco is operating as a general-purpose follow-up facility and is not a dedicated single-use case survey facility. 
The typical performance parameters of the QHY600 cameras are given in Table \ref{tab:sysoverview}. 

The full detector readout with 9k$\times$6k pixels produces large files with a field of view of 1.9$\times$1.2 degrees. 
For most time domain applications a smaller field of view will suffice to observe comparison stars in addition to the main target, and a region of interest readout of the central 30x30 arc minutes reduced field of view mode was introduced to limit the data rate.

The CMOS sensor of the \qhy camera is read out with a rolling shutter.
We will describe the measurement of the readout timing in  Appendix \ref{sec:timing}.

\begin{table}
\renewcommand{\arraystretch}{1.3}
\centering
\begin{tabular}{ll} \hline 
\multicolumn{2}{c}{\lco \deltarho \& QHY600 Properties}
\\ \hline
Telescope     & \planewave \deltarho, 350mm aperture, f/3 \\
Detector size & 	9576 x 6388 pixels, 3.76 micron pixels size; $36 \times 24$ mm in total\\
Field of View & 1.9 x 1.2 degrees readout mode {\tt full\_frame} \\
              & 30'x30' readout mode {\tt central30x30} \\
Projected Pixel Size & 0.73" \\
Cycle Time & $\leq 6$ sec, bias to bias full frame , $\leq$ 3 sec central 30' mode\\
Readnoise & 3.1 to 3.5 \electron typical \\
Gain & 0.7 - 0.8 \electron /ADU \\
Full well & 64kADU $\simeq$ 44k\electron (16-bit limited)\\ 
Filters & Sloan u', g', r', i', z$_\mathrm{s}$, Johnson-Cousins B, V, Pan-STARRS w,\\
        & Baader Planetarium H$_\alpha$, O[III], S[II], \\
        & Astrodon exoplanet (500nm redpass) \\  \hline 
            
\end{tabular}
\vspace{1ex}
\caption{ \label{tab:sysoverview} Key properties of the \qhy camera on the Planewave \deltarho telescope system.}
\end{table}

With up to one year of operating time as of this writing, the \qhy cameras have been working overall very reliably, apart from the occasional need to power cycle the camera due to all 0 readouts or stalled communications. 
Early on we identified some limitations in the humidity management and temperature control systems that we describe below.

\subsubsection{Humidity management in the detector chamber}

An operational issue of all cooled cameras is to avoid condensation or icing in the detector chamber, and we find this especially burdensome in the \lco telescope network where we operate more than 50 cameras on 7 globally distributed sites with limited access to local personnel.
Extending service intervals is paramount in \lco's operational model. 
QHY recommends Silica Gel to be used as a desiccant, and a small container is provided to be attached to the detector chamber. 
In our testing, it became clear that Silica Gel was not the best choice and would result in only about 3-4 months until the set-in of saturation of the desiccant and hence condensation in the camera. 
During further experiments, we found that operating the camera at a cold ($<-10$~deg C) temperature for a prolonged time would lead to ice buildup around the detector while the Silica Gel would not saturate as evidenced by its indicator color.
We suspect that the cold head was effectively regenerating the Silica Gel, and we have moved to Zeolite as a desiccant instead. 

As an additional measure to limit humidity ingress into the detector chamber, we replaced the silicon O-rings (silicon is highly permeable to water vapor) in the camera with Viton O-rings. 

The problem of icing is side-stepped by operating the cameras at a temperature above $0^\circ$C, typically $1^\circ$C or $5^\circ$C, as the dark current remains sufficiently low at these temperatures and is not a dominant noise source in \lco's use cases.  

The QHY cameras have operated for more than 12 months on a single default desiccant canister filled with Zeolite. The QHY600 cameras have humidity sensors built into the detector chamber, which are monitored and linked into the observatory's telemetry system for monitoring and automatic alerting should the chamber humidity rise above a certain threshold.
At such a time, the desiccant will be replaced by on-site personnel.

\subsubsection{Limited range in detector temperature control}

During operation on site, we found detector temperatures below the set point when the ambient temperature dipped below 4 to 5 degrees above the set point. 
As confirmed by QHY, the QHY600 cameras will maintain at least a 5-degree temperature differential to the environment, and offer no capability to counter-heat the detector in cold ambient temperatures. 
This limits the effective dynamical range to a negative 5 to 20 deg temperature differential w.r.t. the environment. 
Given even daily large temperature swings at some sites (e.g., Texas), the detector temperature at which super calibration such as darks are taken at daytime can deviate from the set point reached at night. 
As a mitigation, we resort to two strategies: (i) we seasonally adjust the set point temperature to an achievable level and (ii) to accommodate for intra-day variations, when present, we scale the super dark calibration product to the as-observed science frame detector temperature. 
The limited dynamical range of the \qhy cameras' thermal control system is their most impactful limitation for use in an automated environment.

\subsection {LCO filter wheel and custom motion controller}

\begin{figure}
    \centering
     \includegraphics[width=0.49\textwidth, ]{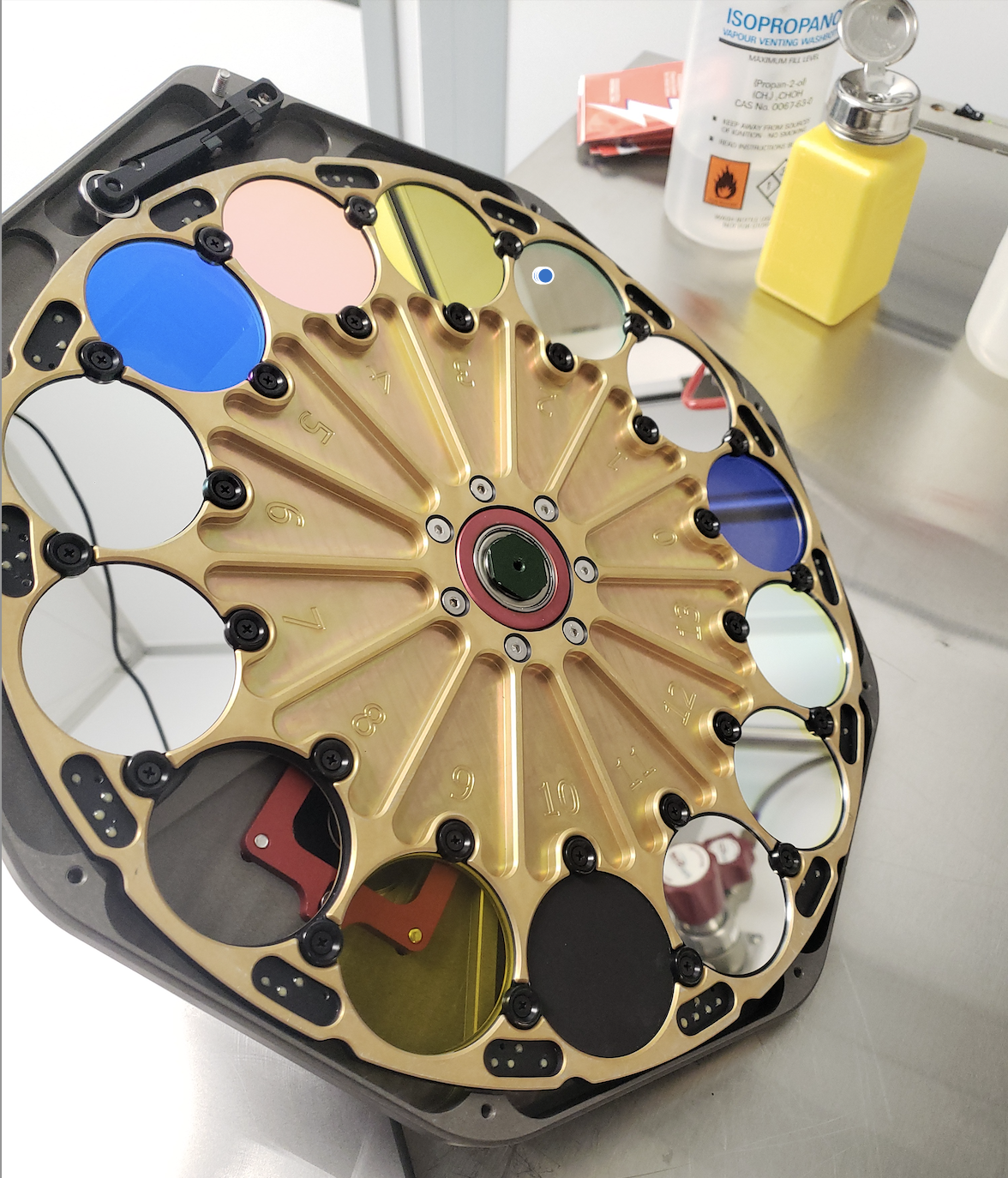}
    \includegraphics[width=0.49\textwidth]{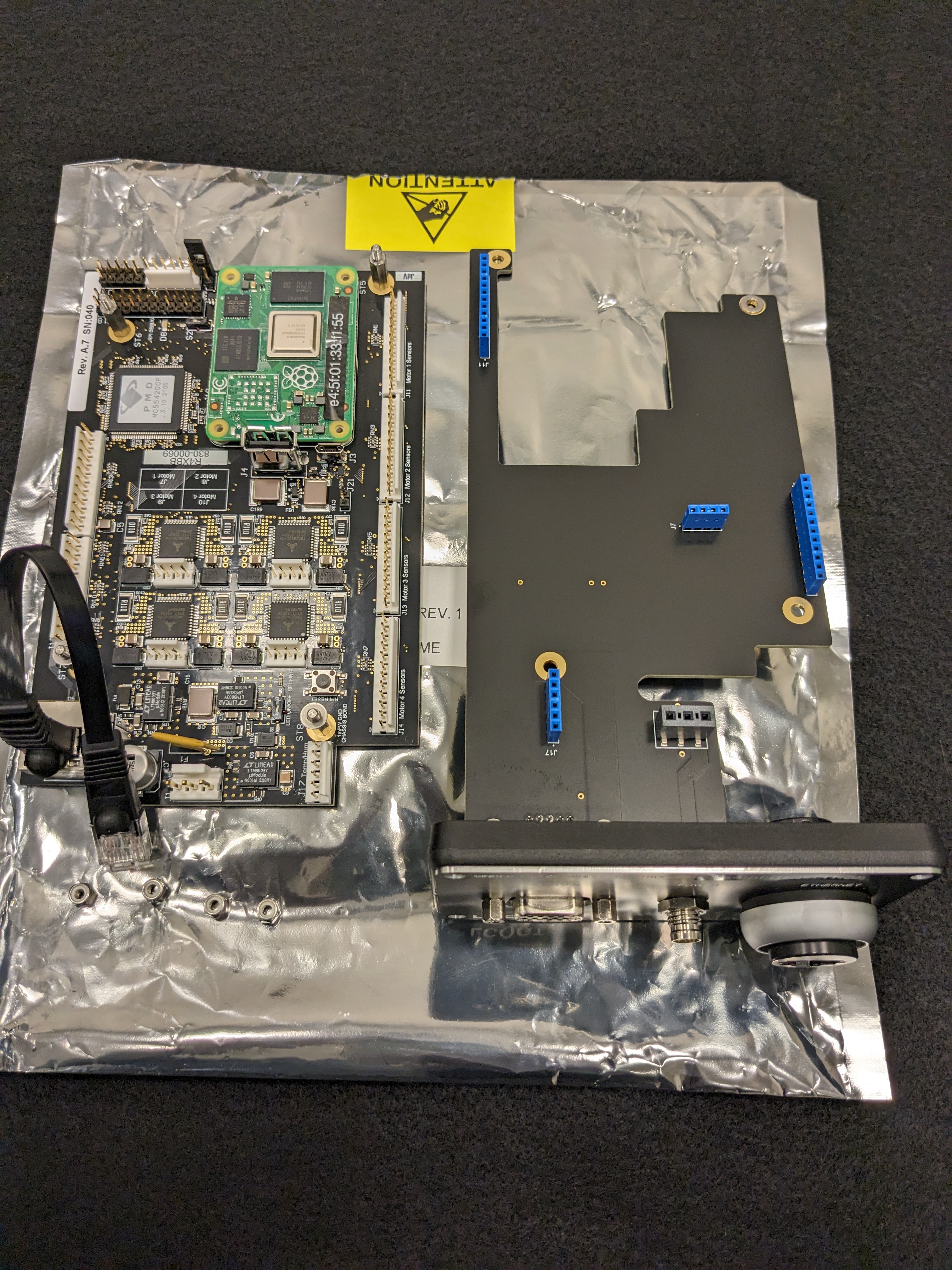}
    \vspace{1ex}
    \caption{\lco 0m4 filter wheel with new filters installed (left) and a custom build 4-channel stepper motor control board ("R4XBB") (right). 
    User interface, API, and high-level programming are provided by a Raspberry Pi 4 compute module. 
    Low-level real-time clocking functions are delegated to an FPGA. 
    For the \deltarho filter wheel, only one of the four channels is utilized. }
    \label{fig:r4xbb}
\end{figure}

The field tested and vetted \lco filterwheels from the previously deployed 0.4-meter network were chosen to accompany the upgraded instrument package. Each filter wheel was returned from the observatories, refurbished, retrofitted with an arrangement of filters as identified in Table \ref{tab:sysoverview}, and optically baffled for the focal ratio of the optical tube assembly (OTA) and instrument package (Figure \ref{fig:r4xbb}, left). 
System commissioning and endurance testing were completed prior to redeployment.

The motion controller for the filter wheel is a new in-house development of a generic 4-channel stepper motor controller board (called R4XBB - Raspberry Pi 4x Base Board).
The core of this control board is an FPGA chip for real-time clocking signal generation and a Raspberry Pi computer module (CM4) for higher-level functions, including hosting an API and providing a web-based user interface.
The R4XBB board was first developed for the 0.4-meter filter wheel application, but with adapted firmware and custom electro-mechanical front-end boards, it is also serving as a motion controller in various components of the \lco 1-meter telescope fleet including driving focus units, mutli-wheel filter exchange mechanisms, and photometric shutters. 
In total, about fifty of those controller boards are in active or planned operation at \lco.

\section{CMOS-specific data processing}

The CMOS detectors in the \qhy600 cameras introduced a new detector technology into the \lco telescope network.
There are two notable differences between CCD and CMOS detectors:

\begin{enumerate}
    \item 
    Each pixel has its own amplifier and hence its own readnoise. 
    The single readnoise used to describe a camera corresponds to the mode of a noise distribution. 
    
    \item 
    The MOSFET transistors in the pixels are susceptible to Random Telegraph Noise (RTN), where in a correlated double-sampling readout of those cameras a pixel could randomly respond with three (or possibly more, but also sometimes only 2!) return values, centered around the expectation value $\pm$ a certain amount of electrons; in the case of the QHY600 cameras up to 80 \electron for the worst pixels on a given detector. 
    An excellent description of RTN in CMOS sensors, including a theoretical foundation, can be found in Wang et al (2006)\cite{wang_random_2006}.
    For an overview of typical RTN noise distribution in the \qhy cameras we refer to Alarcon et al (2023)\cite{alarcon_scientific_2023}.
\end{enumerate}

CMOS cameras in professional astronomy are fairly new, and there is no clear consensus yet on how telegraph noise is to be treated in data processing pipelines. 
As a first step, we have decided to include a limited model of the per-pixel noise behavior in \lco CMOS images so advanced users can improve the noise model of their photometry. 

All \lco imaging data are processed by a standard data pipeline, called BANZAI\cite{mccully_real-time_2018}.
Noise propagation is a key element in the pipeline's design throughout the data processing, starting at individual calibration frames and resulting in a final noise map for the processed image.
The noise map consists of the propagated read noise and shot noise from bias, dark, flat, and science exposures. 
For CCD cameras, this noise map of an individual raw image has been seeded with a single read noise value for the entire image. 
For CMOS cameras, this noise map is seeded with a per-pixel (Gaussian) noise value.

For each detector, this per-pixel CMOS noise map is generated from a set of 50 bias images as part of the commissioning process.
For pixels affected. e.g., by a trimodal RTN distribution, this  Gaussian model is inaccurate and underestimates the peak to valley distribution by a factor of $\sqrt{2}$, but still represents a good starting point to treat RTN at all.

A more complete treatment of RTN would include a per-pixel RTN distribution model and propagation of the RTN model throughout the processing pipeline.
An outline of how such a treatment might look is given in an upcoming paper (Harbeck et al 2024\cite{harbeck_propagating_2024}).

\section {System performance and remaining issues}

\subsection{Throughput}

The QHY600 camera has improved sensitivity in the blue wavelengths compared to the old SBIG 6303 cameras, but CMOS technology at this time still suffers from a reduced Quantum Efficiency (QE) in the near-infrared end of the optical spectrum. 
The collecting area of the \deltarho telescope --- including the central obscuration --- is smaller than the 0.4m Meade telescope by a factor of

$$ \bigl( (0.35/2)^2-(0.196/2)^2 \bigr) / \bigl( (0.406/2)^2 - (0.127/2)^2 \bigr) = 0.56 $$

A comparison of the measured photometric zeropoints between the 0.4-meter Meade and \deltarho systems is shown in Table \ref{tab_photzp}. 
The photometric zero points are given in units of log(\electron/s) and are measured by comparing instrumental magnitudes with local secondary reference stars as listed in the Atlas refcat2 (Tonry et al 2018\cite{tonry_atlas_2018}, Harbeck et al (2022)\cite{harbeck_informing_2022}. 
For all but the z$_{\rm s}$-band filter the increased sensitivity of the new QHY600 camera compensates for the smaller collecting area. 
In combination with a lower readnoise of typically 3.1 \electron versus the previous 14.5 \electron of the SBIG cameras, there is a significant gain in the effective sensitivity of the new system. 

\begin{table}
\renewcommand{\arraystretch}{1.3}
\centering
 \begin{tabular}{lcccc}  \hline 
 filter & zp SBIG6303 kb82 & zp QHY600 sq30 & $\Delta$(mag)& Rel. Sensitivity \\ \hline 
   u' & 16.11 & 17.5 & +1.4 & 3.6 \\
   g' & 21.4  & 21.8 & +0.5 & 1.6 \\
   r' & 21.5 & 21.2 & -0.3 & 0.75 \\
   i' & 20.75 & 20.1 & -0.65 & 0.55 \\
   z$_{\rm s}$ & 19.4 & 18.4 & -1 & 0.4 \\ \hline  
 \end{tabular}
 \vspace{1ex}
 \caption{ Photometric zeropoints (zp) in SDSS filters of the QHY600 + \planewave DeltaRho 0.35m telescope in comparison to the the old SBIG6303 and Meade 0.4m telescope. Note the strong improvement in sensitivity in the u' and g' filters, and the strong decrease in sensitivity in the i' and z' bands.} 
 \label{tab_photzp}
\end{table}

\subsection{Wide field distortions and correction}
\begin{figure}
    \centering
    \includegraphics[width=0.4\textwidth]{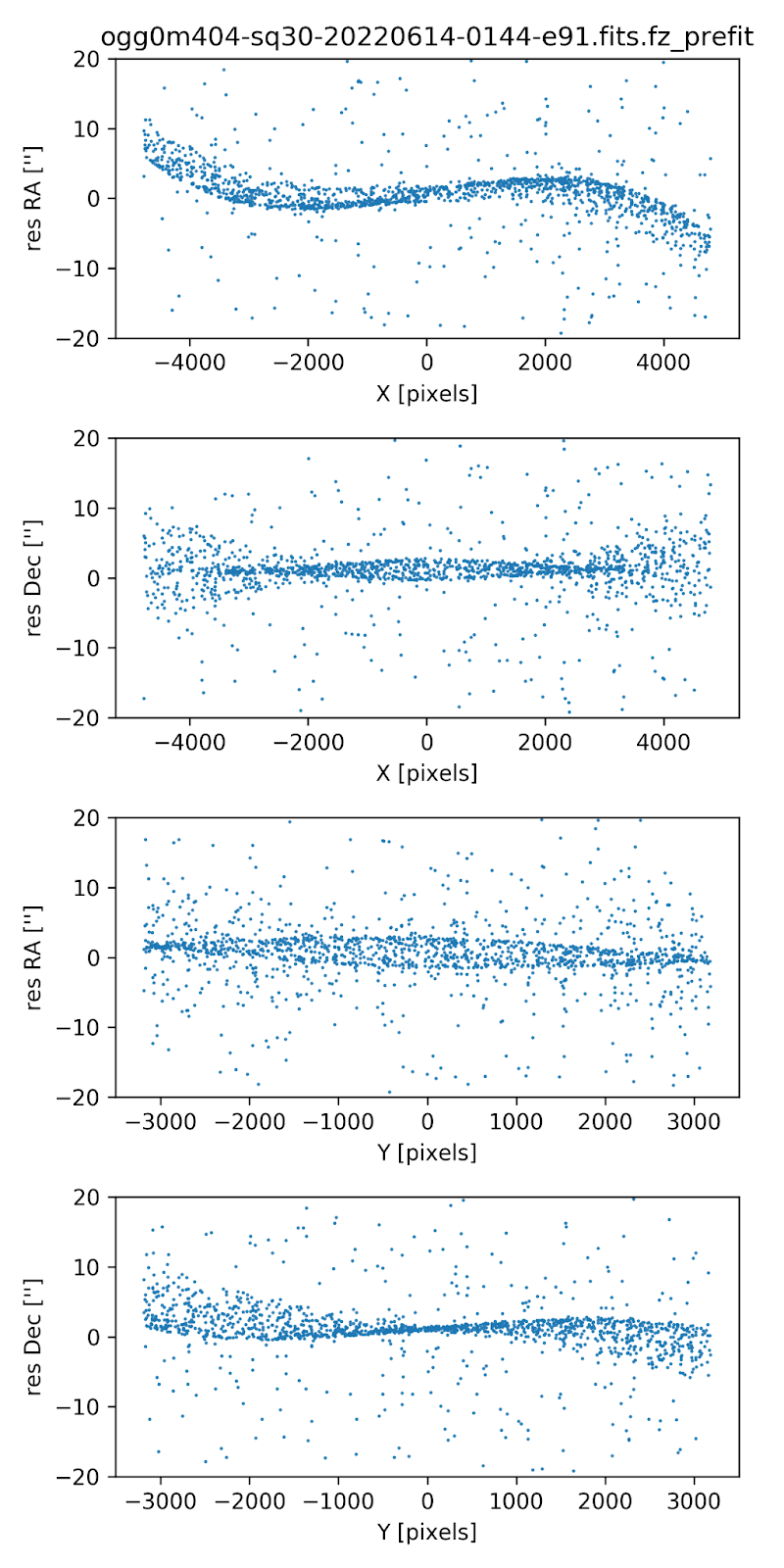}
    \includegraphics[width=0.4\textwidth]{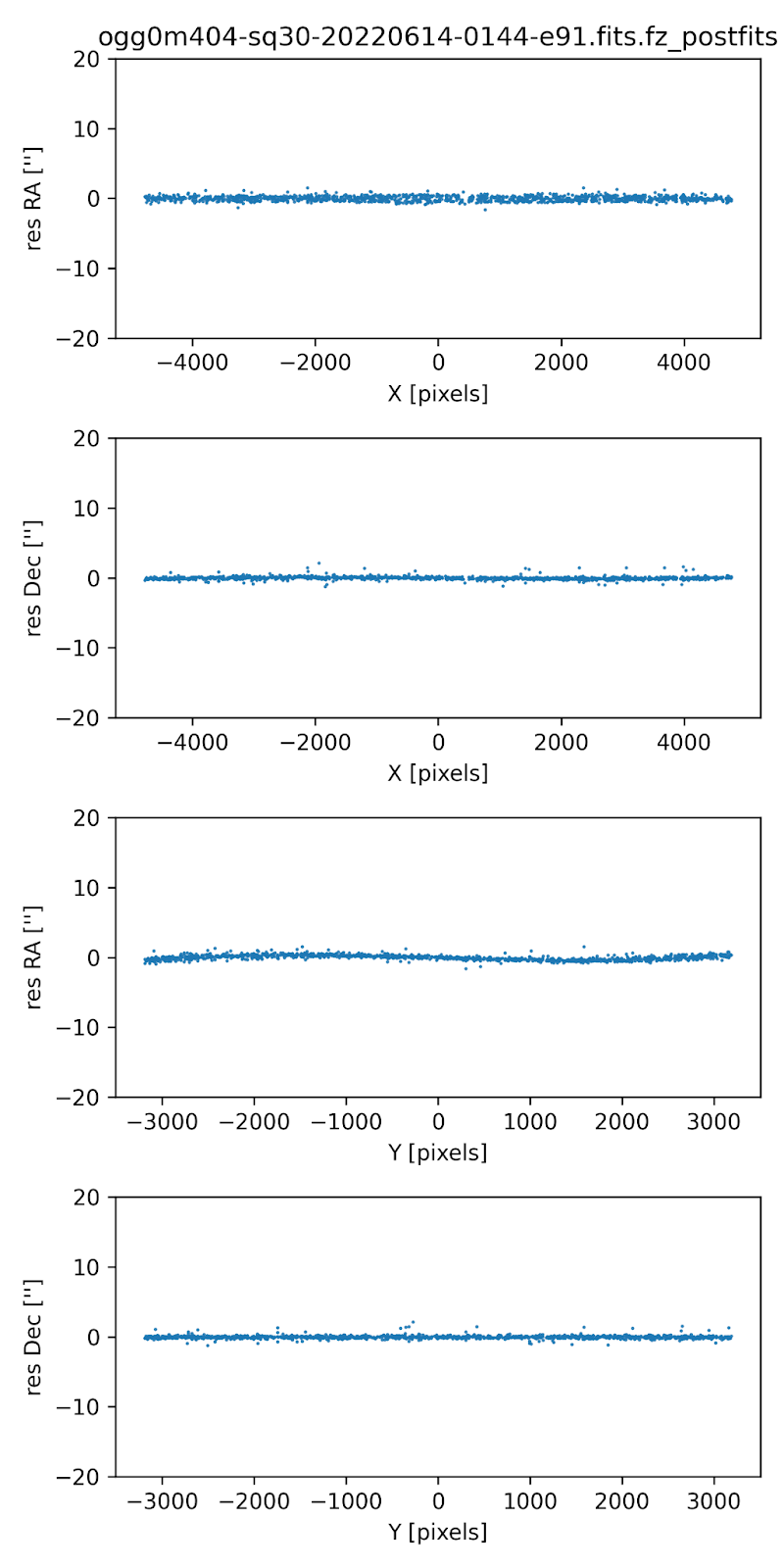}\\
    \vspace{1ex}
    \caption{Stellar position residuals with a linear WCS fit (left) and a third order polynomial correction (right). }
    \label{fig:sip}
\end{figure}

As a wide-field, fast telescope, the DeltaRho is affected by image distortion. 
A linear world coordinate system (WCS) of the full frame QHY600 images yields residuals in the corners on the order of 10 to 15 arcseconds. 
Adding a 3rd order polynomial (SIP) correction will compensate these residuals significantly.
Figure~\ref{fig:sip} shows the improvements in the residuals in R.A. and Dec (differences between star positions and the WCS-modeled positions) from when the image is fit with a linear WCS solution (left column) to when it is fit with a 3rd-order-polynomial-corrected WCS (right column).

\subsection{Tracking, Wind Shake, and Delivered Image Quality}

Limitations to tracking stability are imposed by the telescopes’ C-ring mount.
Unguided images exhibit a drift of up to 20 arcseconds over a few hours. 
Occasionally, the drift occurs within a few minutes, and there can be wind-induced degradation of the image quality. 
The wind limit for scheduling observations at the \deltarho telescopes has been reduced to $9$~m/s from the default observatory limit of $17$~m/s to avoid observing unusable data. 

The low read-noise of the \qhy cameras offers the possibility to break longer exposure requests into multiple shorter exposures to be stacked in software. 
To limit the data volume produced in this process, we are developing an observing mode where the individual shorter exposures will be used to self-guide the telescope but then stacked in realtime in the instrument control software, transparent to the end user. 
Only the final stacked exposure will be preserved, and the individual images will be discarded.

\section{Conclusion}

Over the course of the last year, we have completed the upgrade of the \lco 0.4-meter  education network (now technically a 0.35-meter network, but we preserved the name) at all six globally distributed observatory sites. 
The education and outreach program is a cornerstone of \lco's mission, and the newly upgraded telescope fleet improves the quality of the observing capabilities we can offer through the GSP program: 

The new \qhy cameras significantly improve the photometric accuracy obtainable in this network as evidenced from stellar transit measurements.
The vastly larger field of view in conjunction with the new narrow-band filters enables inspiring astrophotography opportunities.

A core motivation for the network upgrade was to increase the dependability of the 0.4-meter network, especially due to repeated failures with collimation and focus stability. 
The new \deltarho network successfully addressed those core issues.
Thanks to the increased reliability during operations it is possible for \lco to expand the outreach programs on the network as more usable and useful time has become available.

\acknowledgments
 We thank the Gordon and Betty Moore Foundation for generous support for the network upgrade and the Global Sky Partners program.

\appendix

\section{Timing of exposure}
\label{sec:timing}

The \qhy camera facilitates an electronic rolling shutter. The LCO instrument agent software logs the time when an exposure is requested via the QHY SDK.
This timestamp is captured in the \fits file header {\tt UTSTART}.
We expect a certain delay between issuing the expose command and the actual start of the exposure on the detector. 
The electronic rolling shutter of the CMOS chip also introduces a row-dependent delay, and there will be a row-by-row gradient of the effective time of exposure, whereas the integration time remains the same for each row. 
We measure the delay between {\tt UTSTART} and the actual collection of light per row by comparing the exposure level of short exposures illuminated by a GPS-synchronized blinking LED following the description of Marzo et. al. (2021) \cite{marzo_new_2021}.

An Adafruit Ultimate GPS v3 with a PA1616D uBlock GPS receiver  is programmed by a Raspberry Pi Pico controller to blink at 1Hz at a 50\% duty cycle, driven by the GPS module's pulse per second PPS signal.
Once the GPS receiver is locked, the LED PPS will turn on for 0.5 seconds where the rising flank is synchronized to absolute GPS time within $\pm20$ ns jitter according to the PA1616D data sheet. 
The acquisition computer's time is synchronized to a local Stratum 1 GPS network time protocol (ntp) and is typically accurate to a few milliseconds. 

To measure the electronic shuttering effect, we take on the order of fifty 0.5 second long exposures with the QHY camera and plot the fractional seconds of the reported {\tt UTSTART} time versus the received light level on a detector region. 
If the {\tt UTSTART} signal was perfect, exposures (randomly) starting exactly on a full second, when the LED PPS signal turns on would see the maximum light level. 
Exposures starting at a fractional time of 0.5 seconds, when the LED just turned off, would see no light. 
Hence, a maximum light level would be expected when the fractional time was 0 or 1 seconds, and a minimum at 0.5 seconds.
Any delay from the reported {\tt UTSTART} time of the exposure to the true exposure would shift the minimum to the left; 
an exposure earlier than the reported {\tt UTSTART} would shift the relation to the right. 

\begin{figure}
    \centering
    \includegraphics[width=0.49\textwidth]{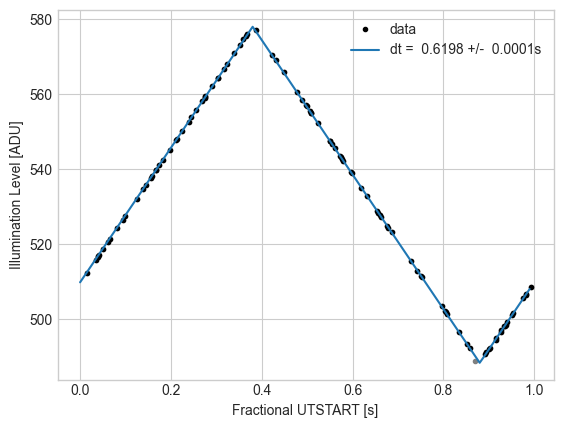}
    \includegraphics[width=0.49\textwidth]{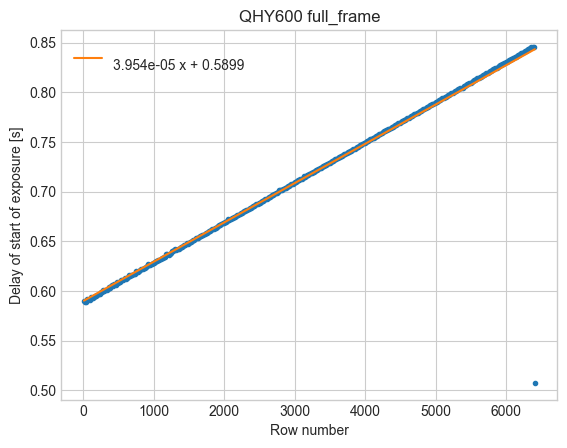}\\
    \caption{Left: Fractional reported UTSTART time of exposure vs observed light level of a select region on the QHY600 camera. A phase-shifted (dt) triangular function is fitted to the data to determine the delay between true start of exposure and the assumed start of exposure.  Right: The fitted exposure delay dt for the rows of a full-frame readout, illustrating the rolling shutter readout mode.  Note that the absolute offset and time per row are identical regardless of windowing. }
    \label{fig:gpslock}
\end{figure}

For the average light level of each detector row,  we fit a triangular function to the measured exposure level, where the phase shift dt (constraint from [0..1) seconds), the amplitude and bias level are free  parameters, as exemplified for one select row in  (Figure \ref{fig:gpslock} left).
In the same Figure (right) we plot the row number versus the measured readout delay dt for that row and fit a linear function to describe the total offset each row would encounter during the rolling shutter readout.

With this linear correlation, the corrected start of exposure UTSTART$_C$ is for a given row number (row\#) is: 
$$ \rm{UTSTART}_C = \rm{UTSTART} + 0.5899s + 39.54\cdot 10^{-6} s/row \times row\#$$

E.g., for the full frame readout, the center of the image is row \#3194, and the effective exposure delay dt is:

$$ dt =  0.5899s + 39.54\cdot 10^{-6} s/\rm{row} * 3194~ \rm{rows} = 0.7162 s$$

The relation (offset and per row time) is the same for both the full\_frame and central30x30 readout modes and applies to the row number of the respective readout mode.

\bibliography{references} 
\bibliographystyle{spiebib} 

\end{document}